\documentclass[aps,twocolumn,showpacs,nofootinbib,amssymb]{revtex4}
\usepackage{graphicx,epsfig}
\newcommand{\lp}{\left(}
\newcommand{\rp}{\right)}

\newcommand{\nn}{\nonumber}
\newcommand{\be}{\begin{equation}}
\newcommand{\ee}{\end{equation}}
\newcommand{\bea}{\begin{eqnarray}}
\newcommand{\eea}{\end{eqnarray}}

\def\siml{{\ \lower-1.2pt\vbox{\hbox{\rlap{$<$}\lower6pt\vbox{\hbox{$\sim$}}}}\ }}

\begin{document}
\preprint{UAB-FT-XXX}

\title{\boldmath 
Breakdown of the operator product expansion in the 't~Hooft model
}
\author{Jorge Mondejar$^1$ and Antonio Pineda$^2$}
\affiliation{$^1$Dept.~d'Estructura~i~Constituents~de~la~Mat\`eria,~U.~Barcelona,
~E-08028~Barcelona,~Spain\\
$^2$ Grup de F\'\i sica Te\`orica and IFAE, Universitat
Aut\`onoma de Barcelona, E-08193 Bellaterra, Barcelona, Spain}

\date{\today}

\begin{abstract}
\noindent
We consider deep inelastic scattering in the 't~Hooft model. Being solvable, 
this model allows us to directly compute the moments associated to the cross section 
at next-to-leading order in the 
$1/Q^2$ expansion. We perform the same computation using the operator product expansion. 
We find that all the terms match in both computations except for one in the hadronic side, which 
is proportional to a non-local operator.  
The basics of the result suggest that a similar phenomenon may occur in four dimensions in 
the large $N_c$ limit.
\end{abstract}
\maketitle
It has been more than 35 years since QCD was vindicated as the theory of the 
strong interactions \cite{Gross:1973id}. Unfortunately, analytic solutions describing 
the hadrons and their interactions 
in terms of the degrees of freedom (quarks and gluons) and parameters (coupling constant and quark masses) 
that appear in the QCD Lagrangian have remained elusive. Leaving aside symmetry considerations (or how they are realized), 
the only quantitative and analytic scheme to check the dynamics of QCD from 
first principles is through (perturbative) weak-coupling computations. The problem then 
is to relate those computations to experiments, which are naturally described in terms of hadrons. 
In order to do so two key ingredients are used:

1) Asymptotic freedom \cite{Gross:1973id,Politzer:1973fx}, which states that the renormalized coupling goes to zero at 
large Euclidean momentum. 

2) The operator product expansion (OPE) \cite{Wilson:1969zs} and, more generally, factorization. 
So far, the OPE has only been proven within perturbation theory \cite{Zimmermann:1972tv}.  

In practice, the combination of these two points is stated as the fact that the multiplication of two operators 
(sandwiched between physical states) enjoys the following expansion at short distances 
\be
\hat A(x)\hat B(0)=\sum_nC_n(x)\hat O_n(0)
\,,
\ee
where $\hat O_n$ are local operators with increasing dimensionality in $n$ 
and the coefficients $C_n$ can be computed in perturbation theory (or at least at weak coupling). 

The (Fourier transform of the) 
OPE probes these correlators at large Euclidean momentum. Therefore, the OPE is not directly accessible to experiment 
and one has to resort to dispersion relations\footnote{These may have potential problems of their own; basically 
the lack of knowledge of the asymptotic behavior of the correlators to ensure that one can neglect the contributions 
at infinity when using the Cauchy theorem. This will not affect the main conclusions of this paper, since this 
ambiguity should have an impact on, at most, a finite number of moments.} to check it. 
This practical version of the OPE is at the basis of computations at large Euclidean momentum 
of (the moments in) deep inelastic scattering (DIS) and the vacuum polarization tensor, which so far have been 
thought to be among the more solid predictions of QCD, since they are not affected by quark-hadron duality 
problems\footnote{We do not enter in this paper on the use of the OPE 
and factorization methods for quantities living in, or affected by, the Minkowski cut. These quantities 
are usually regarded as less fundamental, and are the ground on which 
the discussion about quark-hadron duality takes place, see for instance \cite{Shifman:2000jv} and references therein. 
For those observables one can easily find examples where perturbation theory fails 
in the large $N_c$ limit. For instance, the imaginary part of the vacuum polarization tensor becomes 
a sum of infinitely narrow resonances in the large $N_c$ limit, as opposed to the smooth result obtained 
from perturbation theory.}. 
Therefore, the importance of setting the OPE and the factorization methods used in quantum field theories, 
specially in QCD, on solid theoretical ground can hardly be overemphasized. 
So far, it was thought that the use of the OPE (in its non-perturbative formulation \cite{OPE}) was secure, 
even tough it has not been proven in QCD. It has been only partially checked in models, for 
instance in two dimensional QCD in the large $N_c$ limit: 
the 't Hooft model \cite{hooft2}. This theory is superrenormalizable and asymptotically free, 
so it is a nice ground on which to test 
the OPE\footnote{In the 't Hooft model there are no marginal operators. Therefore, 
the coupling constant does not run and has dimensions; 
no renormalons should then arise.}. 
This was done at the lowest order in the OPE in Refs. \cite{callan,einhorn} for the vacuum polarization and for  
DIS off a meson with nice agreement between the results of the model and the OPE expectations. 
In Ref. \cite{Mondejar:2008dt} the OPE was numerically checked at 
next-to-leading order (NLO) in the $1/Q^2$ expansion, with logarithmic accuracy, for the vacuum polarization. 
In this paper we consider DIS at NLO. We will only give the main results and postpone the 
details to a later publication. In practice, we will compute the correlator
\bea
\label{Tmunu}
&&T^{\mu\nu}(q)=i\int d^2xe^{iq\cdot x}
\langle ij;n|T
j^{\mu}(x)j^{\nu}(0)
|ij;n\rangle \\
&&\equiv
\lp P_n^{\mu}-\frac{q^{\mu}q\cdot P_n}{q^2}\rp
\lp P_n^{\mu}-\frac{q^{\mu}q\cdot P_n}{q^2}\rp
 T(Q^2,x_B) \nn
 \,,
\eea
where $x_B=Q^2/(2P_n\cdot q)$, $Q^2=-q^2$, $P_n$ is the momentum of the meson 
and $i$, $j$ stand for the flavor of the quark and antiquark, respectively, which form the bound state.
$j^{\mu}(x)=\sum_{h}j_h^{\mu}(x)$, where $j_h^{\mu}(x)=e_h\bar{\psi}_h\gamma^{\mu}\psi_h(x)$. 
  
The imaginary part of $T^{\mu\nu}$ is proportional to the differential cross section when 
$x_B\geq 0$, and in the light-cone frame reads  
\bea
\nn
{\rm Im}T^{++}&=&\frac{1}{2}
\sum_m\int\frac{dP_m^+}{2(2\pi)P_m^+}
\left|\langle ij; m|j^{+}(0)
|ij; n\rangle\right|^2
\\
\nn
&&
\times
(2\pi)^2\delta^2(q+P_n-P_m)
\\
\nn
&=&
\pi
\sum_m
\left|\langle ij; m|\sum_{h=i,j}e_hj_h^{+}(0)|ij; n\rangle\right|^2 
\\
\label{ImT++}
&&
\times
\delta\lp M_m^2-M_n^2-Q^2\frac{(1-x_B)}{x_B}\rp 
\ .
\eea
Using dispersions relations, 
up to a finite number of subtractions, one should have (see also the discussion in Ref. \cite{Burkardt:1991hu})
\bea
\label{TCauchy}
T(Q^2,x_B)
=
\frac{2}{\pi}
\int_0^{x_B^{max}}d y_B\frac{1}{y_B}\frac{{\rm Im}{T}(Q^2,y_B)}{1-\lp \frac{y_B}{x_B}\rp^2-i\epsilon}
\,,
\eea
where $x_B^{max}=1/(1+(P_0^2-P_n^2)/Q^2)$. Therefore, 
$T(Q^2,x_B)$ admits an analytic expansion in $1/x_B$ for $x_B > x_B^{max}$,
\be
\label{TMN}
 T(Q^2,x_B)=
4\sum_{N=0,2,4,...} M_N(Q^2)\frac{1}{x_B^N} \ ,
\ee
where
\bea
\label{MNdef}
M_N(Q^2)\equiv \frac{1}{2\pi}\int_0^{x_{B}^{max}}dy_B y_B^{N-1}{\rm Im}{T}(Q^2,y_B) 
\,.
\eea
General expressions for the matrix elements in terms of the 't Hooft wave functions, 
$\phi_n^{ij}(x)$, will be presented in Ref. \cite{prep}. They are obtained using 
similar techniques to those in Ref. \cite{Mondejar:2006ct}, 
where the Lagrangian, the 't Hooft equation, as well as the definitions of the bound states 
are given. Approximated expressions, valid when $1-x_B \gg \beta^2/Q^2$ 
($\beta^2=g^2N_c/(2\pi)$ is the 't~Hooft coupling), 
will also be obtained in Ref. \cite{prep} using Eqs. (51) and (52) from \cite{Mondejar:2008dt}, as well as similar 
techniques to those used in Ref. \cite{Mondejar:2006ct}. 
For the explicit computation it is convenient to work with the kinematical variable 
$x=-q^+/P_n^+$ (we take $q^+ <0$ and $q^- \rightarrow +\infty$), 
which satisfies the equality $x_B=x/(1-\frac{M_n^2}{Q^2}x^2)$.  
One then obtains Im$T$ with ${\cal O}(1/Q^2)$ precision 
when $1-x \gg \beta^2/Q^2$
\bea
\label{ImThad}
&&{\rm Im}T
\simeq
4\pi\left(\frac{2\pi\beta x}{Q^2}\right)^2\frac{1}{\left(1+x^2\frac{M_n^2}{Q^2}\right)^2}
\\
\nn
&&
\times
\sum_{m=0}^{\infty} \delta\lp M_m^2-M_n^2(1-x)-Q^2\frac{(1-x)}{x}\rp 
\\
\nn
&&
\times
\left[
e_im_i\left\{
\phi^{ij}_n(x)
\lp 
1-\frac{m_i^2+m_im_j(-1)^{m}}{Q^2}
\right.
\right.
\right.
\\
\nn
&&
\left.
\left.
+
\left(
\frac{m_{i,R}^2+m_{j,R}^2}{2Q^2}
+
\frac{m_{i}m_{j}}{Q^2}(-1)^{m}
\right)
\frac{x}{1-x}
\rp
\right.
\\
\nn
&&
\left.
\left.
+\frac{x}{Q^2}(m_{i,R}^2+m_im_j(-1)^{m})
\frac{d\phi^{ij}_n(x)}{dx}\rp
\right\}
\\
\nn
&&
\left.
-(-1)^{m}
		 e_jm_j 
		 \left\{
\phi^{ij}_n(1-x)
\lp 
1-\frac{m^2_j+m_jm_i(-1)^{m}}{Q^2}
\right.
\right.
\right.
\\
\nn
&&
\left.
\left.
+\left(
\frac{m_{j,R}^2+m_{i,R}^2}{2Q^2}
+
\frac{m_{j}m_{i}}{Q^2}(-1)^{m}
\right)
\frac{x}{1-x}
\rp
\right.
\\
\nn
&&
\left.
\left.
+\frac{x}{Q^2}(m_{j,R}^2+m_jm_i(-1)^{m})\frac{d\phi^{ij}_n(1-x)}{dx}\rp
\right\} 
\\
\nn
&&
\qquad
\left.
+
o\left(\frac{1}{Q^2}\right)\right]^2
\ ,
\eea
where $m_{i,R}^2=m_i^2-\beta^2$. Up to a prefactor, the terms 
$e_{i/j}m_{i/j}\left\{\cdots\right\}$ represent the contribution from the matrix element of 
the current of the quark $i$ and the antiquark $j$, respectively. Note the relative sign 
$-(-1)^m$ between 
both contributions, which can be obtained from symmetry arguments. 

Eq. (\ref{ImThad}) is one of the key results of our paper. 
By inserting this expression in the moments and using the Euler-MacLaurin expansion we obtain 
($x_{max}= 1-M_0^2/Q^2+{\cal O}(1/Q^4)$) 
\bea
\nn
&&
M_N(Q^2)
=
\frac{8}{Q^4}
\int_0^{x_{max}}dx
\left(\frac{x}{1-\frac{M_n^2}{Q^2}x^2}\right)^N\frac{x}{1-\frac{M_n^4}{Q^4}x^4}
\\
\nn
&&
\times
\left[
e_i^2m_i^2\left(\phi_n^{ij}(x)\right)^2
+
e_j^2m_j^2\left(\phi_n^{ij}(1-x)\right)^2
\right.
\\
\nn
&&
+
2e_i^2m_i^2\phi_n^{ij}(x)\left[-\frac{m_i^2}{Q^2}\phi_n^{ij}(x)+x\frac{m_{i,R}^2}{Q^2}\frac{d\phi_n^{ij}(x)}{dx}\right]
\\
\nn
&&
+
2e_j^2m_j^2\phi_n^{ij}(1-x)\left[-\frac{m_j^2}{Q^2}\phi_n^{ij}(1-x)
\right.
\\
\nn
&&
\left.
\qquad\qquad
+x\frac{m_{j,R}^2}{Q^2}\frac{d\phi_n^{ij}(1-x)}{dx}\right]
\\
&&
\nn
-
4e_ie_jm_i^2m_j^2\frac{2x-1}{Q^2(1-x)}\phi_n^{ij}(x)\phi_n^{ij}(1-x)
\\
&&
\left.
-
2e_ie_jm_i^2m_j^2\frac{x}{Q^2}\frac{d}{dx}\left(\phi_n^{ij}(x)\phi_n^{ij}(1-x)\right)
\right]
\,,
\label{MN}
\eea
which is correct with ${\cal O}(1/Q^2)$ precision (for finite $N$). Note that with this precision 
we can replace $x_{max}=1$. Note as well that the last two terms in 
Eq. (\ref{MN}) are ${\cal O}(m^2/Q^2)$ suppressed 
with respect to the leading term. 
One comment is in order here. When using the 
Euler-MacLaurin expansion we had to deal with terms proportional to $(-1)^m$. Such terms give a potential 
contribution of relative order $o\left(1/Q^2\right)$, 
beyond the precision of our computation.
Nevertheless, 
we also have terms proportional to $(-1)^{2m}=1$. They come from the interference of the particle and antiparticle currents 
and are no longer suppressed by the sign-alternating behavior, only by 
the prefactor, which is ${\cal O}(1/Q^2)$. Therefore, these terms contribute at ${\cal O}(1/Q^2)$. We anticipate that they will be the ones responsible for the violation of the OPE.  

Using Eq. (\ref{TMN}) and the approximated expressions we have obtained for the moments in Eq. (\ref{MN}), we can obtain an 
approximated expression for $T$, which can actually be written as a dispersion relation formula: 
\be
\label{ImThadr}
T^{Eul.}(Q^2,x_B) =\frac{2}{\pi}\int_{0}^{1}\frac{d y_B}{y_B}\frac{{\rm Im}T^{Eul.}(Q^2,y_B)}{1-\lp \frac{y_B}{x_B}\rp^2-i\epsilon} \ ,
\ee
where $\text{Im}T^{Eul.}$ is given by Eq. (\ref{ImThad}), performing the substitution $\sum_m\to\int dm$.
$T^{Eul.}$ is not a good approximation to $T$ when we approach the physical cut. This has to do with the fact our 
expressions for the moments are not valid when $N \rightarrow \infty$ ($x \rightarrow 1$ limit). Nevertheless, it can be 
considered a generating functional for the moments with not very large $N$. Therefore, it is the expression we would expect 
to be equal to the OPE result. In order to ease the comparison, we write $T^{Eul.}$ in a factorized way:
\bea
\label{TEul}
&&
T^{Eul.}(Q^2,x_B) = 
-2\lp\frac{4}{Q^2}\rp^2
\\
\nn
&&
\times
\int_{-\infty}^{\infty} dy \left\{ e_i^2J_i(x,y) f_i(y)+e_j^2J_j(x,y) f_j(y)\right.
 \\
 \nn
&&\left.+ e_ie_j J_{int.}(x,y)f_{int.}(y)\right\} \ ,\nn
\eea
where  
\bea
\nn
&&f_i(y)\equiv 
\left[\phi_n^{ij}(y)\right]^2
= 
\int_{-\infty}^{\infty} \frac{dx^-}{2(2\pi)}
e^{-iy P_n^+ \frac{x^-}{2}}
\\
\label{fi}
&&
\times
  \langle ij;n| \psi_{i,+}^{\dagger}(x^-)
 \psi_{i,+}(0)|ij;n\rangle
 \,,
 \\
\nn
&&f_j(y)\equiv
\left[\phi_n^{ij}(1-y)\right]^2
= 
-\int_{-\infty}^{\infty}\frac{dx^-}{2(2\pi)}
e^{-iy P_n^+ \frac{x^-}{2}}
\\
&&
\label{fj}
\times
   \langle ij;n| \psi_{j,+}^{\dagger}(0)
  \psi_{j,+}(x^-)|ij;n\rangle
  \,, 
\eea
\bea
\label{finti}
 &&f_{int.}(y)\equiv\frac{m_im_j}{y(1-y)}\phi_n^{ij}(y)\phi_n^{ij}(1-y) 
 \\
\nn
&&
=
\frac{(P_n^+)^2}{N_c}\int_{-\infty}^{\infty} \frac{dx^-}{2(2\pi)} 
e^{-iy P_n^+ \frac{x^-}{2}}\int_{-\infty}^{\infty} dz^- 
\\
\nn
&&
\times
\langle ij;n|\psi^{\dagger}_{i,-}(x^-)\psi_{j,-}(z^-)
\psi^{\dagger}_{j,+}(0)\psi_{i,+}(z^-)
|ij;n\rangle \ .
\eea
In the above expressions we have not inserted the Wilson line:
\be
\label{Wline}
\Phi(x^-,y^-)=P[e^{(ig\int_{y^-}^{x^-}dz^-A^+(z^-))}] \ ,
\ee
between the quark fields $\psi$ to make gauge invariance explicit, since we are working in the light-cone gauge, $A^+=0$. 

The functions $J$ are defined as\footnote{It is possible to redefine $J$ so that it has the functionality 
$J(y/x)$. This implies redefining $f(y)$ by some powers of $y$, equivalent to introducing some extra $\partial^+$ 
derivatives. We do not do so in this paper because it would increase the length of the formulae. Note as well 
that the derivatives in $J_i$ tend to change the variable of $f_i(y)$ to $f_i(y(1+m_{i,R}^2/Q^2))$, consistent with 
the interpretation from the perturbative computation in the OPE.}
\bea
\label{JX}
J_{i(j)}(x,y)
&\equiv&
\frac{m_{i(j)}^2x^2y}{y^2-x^2+i\epsilon}\left[ 1-2\frac{m_{i(j)}^2}{Q^2}
\right.
\\
\nn
&&
\left.
-2\frac{M_n^2}{Q^2}y^2+\frac{m_{i(j),R}^2}{Q^2}y\frac{d}{dy}\right]
\,,
\\
J_{int.}(x,y)&\equiv& 2\frac{m_im_j}{Q^2}\left[x^2\frac{2y^2(1-2y)}{y^2-x^2+i\epsilon}\right.\nn\\
&&\left.-x^3\frac{d}{dx}\frac{y^2(1-y)}{y^2-x^2+i\epsilon}\right] \ .
\eea

We can now perform the same computation using the OPE (we need to do the computation at one loop). 
In order to avoid spurious differences with the hadronic result, we compute Im$T^{OPE}$ 
and use dispersion relations afterwards (actually, one can already see the failure of the OPE 
calculation from the comparison of Im$T^{OPE}$ and Im$T^{Eul}$). We then obtain 
\bea
\label{TOPE}&&
T^{OPE}(Q^2,x_B) = -2\lp\frac{4}{Q^2}\rp^2
\\
&&
\times
\int_{-\infty}^{\infty} dy \left\{ 
e_i^2J_i(x,y) f_i(y) +e_j^2J_j(x,y)f_j(y) \right\}  \nn 
\,.
\eea
We can easily see that $T^{OPE} \not= T^{Eul}$. If we consider the moments generated by 
$T^{OPE}$, $M_N^{OPE}$, 
they can be expressed in terms of non-perturbative local operators (for the explicit expressions see \cite{prep}) 
as expected. Nevertheless, $M_N$ has extra terms. If we consider the difference, we obtain
 \bea
\nn&&M_N(Q^2)-M_N^{OPE}(Q^2)
=\frac{16e_ie_j}{Q^4}\frac{1}{N_c}\frac{m_im_j}{Q^2} \nn \\
&&\times \int dz^- \langle ij;n| \psi_{i,-}^{\dagger}(0) \frac{(-i\overleftarrow{D}^+)^{N+2}}{(P_n^+)^{N+1}}\nn\\
&&\times\left[(N+4)\lp 1-\frac{-i\overleftarrow{D}^+}{P_n^+}\rp-2 \frac{-i\overleftarrow{D}^+}{P_n^+}\right] \nn\\
&&\times\psi_{j,-}(z^-)\psi_{j,+}^{\dagger}(0)
\psi_{i,+}(z^-)|ij;n\rangle \ ,
\label{MNmat}
\eea
which is expressed in terms of non-local operators. 
We take this 
result as evidence of the existence of OPE-violating terms. 
Let us stress that this is the first time, that we are 
aware of, that an analytic calculation in a quantum field theory 
exhibits OPE-violating contributions. 

The possible existence 
of OPE-breaking effects in QCD has already been discussed in the past. 
As early as in Ref. \cite{Burgio:1997hc} numerical evidence for the existence of OPE-breaking 
effects in the gluon condensate was claimed. Nevertheless, it is still 
unclear whether those effects can be associated to ultraviolet renormalons and/or higher 
orders in perturbation theory (for a recent discussion see \cite{Rakow:2005yn}). Over the years 
there has also been some discussion on the possible existence of a $\langle A^2 \rangle_{min.}$ 
condensate. This object should actually correspond to a non-local gauge-invariant condensate, though 
its explicit form is unknown for QCD \cite{Gubarev:2000eu}. Finally, there are some models that 
may produce effects that break the OPE, see for instance \cite{Dorokhov:2003kf}.
Nevertheless, those OPE-breaking effects would affect the static potential and 
the vacuum polarization. Regarding this we would like to emphasize that we do not 
find any OPE-breaking effect in the static potential or the vacuum polarization in 
the 't Hooft model. The static potential can be computed exactly in the 't~Hooft model 
within perturbation theory. Therefore, there is no room there for effects associated to a 
sort of $\langle A^2 \rangle_{min.}$ condensate. With the present precision of our 
computation, we also do not see OPE-breaking effects in the vacuum polarization. 

The existence of this OPE-breaking effect is much associated to the large $N_c$ analysis we have done, 
and the existence of sign-alternating effects. It should not be difficult to devise similar large $N_c$-inspired models in 4 dimensions, 
which would produce OPE-violating terms. In those models, however, it might be difficult to disentangle the OPE-violating terms from standard OPE contributions, 
since they both would scale in the same (or a very similar) way and, at the end of the day, one fits higher twist effects to data. 

The new term is analytic in $1/Q^2$ and can actually be written in a factorized form. Therefore, 
one may think whether one could extend the standard formulation of the OPE in order to include this sort of terms. It would also be 
interesting to get a better understanding of this term from a diagrammatic analysis, if possible. 

In conclusion, we have shown that in the 't~Hooft model, at NLO in the $1/Q^2$ expansion, the 
moments associated to DIS receive a contribution that:\\
1) cannot be written in terms of local operators,\\
2) cannot be matched with any OPE-like contribution we are aware of.\\ 
We take these two facts as ``smoking-gun" signals for the breakdown of the OPE in DIS 
in the 't Hooft model at NLO in the $1/Q^2$ expansion. Note that it appears as a subleading (NLO) effect in DIS. 
Moreover, it is important that we considered the interference between two currents, otherwise this effect would be 
suppressed by a factor of, at least, order $1/Q^4$. For the vacuum polarization we do not see this sort of effects 
with the present precision \cite{Mondejar:2008dt}. 

\medskip

{\bf Acknowledgments}. We thank S. Peris for discussions. This work is partially supported by the 
network Flavianet MRTN-CT-2006-035482, by the spanish 
grant FPA2007-60275, by the Spanish Consolider-Ingenio 2010
Programme CPAN (CSD2007-00042), and by the catalan grant SGR2005-00916.

\end{document}